\documentclass{llncs}

\usepackage[utf8]{inputenc}
\usepackage[para]{footmisc}
\usepackage[table,xcdraw,dvipsnames]{xcolor}
\usepackage[numbers,sort&compress]{natbib}
\usepackage[english]{babel}

\usepackage{amsmath}
\usepackage{amssymb}
\usepackage{enumitem}
\usepackage{etoolbox}
\usepackage{float}
\usepackage{graphicx}
\usepackage{hyperref}
\usepackage{mathrsfs}
\usepackage{multirow}
\usepackage{subcaption}
\usepackage{url}
\usepackage{wrapfig}

\hypersetup{
  colorlinks,
  linkcolor = {red!50!black},
  citecolor = {blue!50!black},
  urlcolor  = {blue!30!black}
}

\begin{document}

\title{Topological street-network characterization through feature-vector and cluster analysis}
\titlerunning{Topological characterization of street networks}

\author{Gabriel Spadon \and
Gabriel Gimenes \and
Jose F. Rodrigues-Jr}
\authorrunning{G. Spadon et al.}

\institute{%
University of Sao Paulo, Sao Carlos -- SP, Brazil\\ \ \\
\email{spadon@usp.br},
\email{ggimenes@icmc.usp.br},
\email{junio@icmc.usp.br}}

\maketitle

\begin{abstract}
Complex networks provide a means to describe cities through their street mesh, expressing characteristics that refer to the structure and organization of an urban zone.
Although other studies have used complex networks to model street meshes, we observed a lack of methods to characterize the relationship between cities by using their topological features.
Accordingly, this paper aims to describe interactions between cities by using vectors of topological features extracted from their street meshes represented as complex networks.
The methodology of this study is based on the use of digital maps.
Over the computational representation of such maps, we extract global complex-network features that embody the characteristics of the cities.
These vectors allow for the use of multidimensional projection and clustering techniques, enabling a similarity-based comparison of the street meshes.
We experiment with 645 cities from the Brazilian state of Sao Paulo.
Our results show how the joint of global features describes urban indicators that are deep-rooted in the network's topology and how they reveal characteristics and similarities among sets of cities that are separated from each other.
\end{abstract}

\section{Introduction and Related Works}
\label{sect:introduction}

Complex networks are used to shape real-world systems, {\it e.g.} networks of protein interaction, street meshes, and subway lines.
These networks, as mathematical models, stand out due to their algebraic properties and computing potential, with analytical applicability to support cognitive processes of decision-making~\cite{Boccaletti2006:StructureDynamics}.
Through metrics and methods based on topology and/or geometry, it is possible to identify characteristics of interest that are not obvious for human inspections based on reading; this is because the networks may be wide (high number of vertices), intricate (high number of edges), or may hold non-trivial patterns and attributes whose observation depends on the application of algorithms.

In the specific case of the representation of street networks, complex networks describe factors related to the displacement of individuals, allocation of services, the improvement of tasks related to transport, and even to the study of factors from collective behavior, when the network is weighted by the associated data.
In this regard, we observed a lack of methods to characterize groups of cities by means of the features that can be extracted from their topology, which is the aim of this research.
This methodology has potential to enhance the understanding of an urban space and to explain the reason why cities share properties of interest.

To this end, we developed a methodology composed of {\it Data Acquisition and Preparation}, {\it Feature Extraction and Selection}, and {\it Feature Vector Analysis}.
We analyzed 645 cities from the state of Sao Paulo, aiming to provide comprehension of peculiarities from different cities by interpreting global network-characteristics.
These cities are representations of street meshes that were extracted from digital maps, such that they were gathered and analyzed by using machine-learning methods of feature extraction, multidimensional projection, and cluster analysis.
In order to demonstrate our methodology, we investigate the following hypotheses:
{\bf (A)} {\it the network topology is a tool-set that can reveal groups of cities with similar characteristics, potentially revealing disparities};
{\bf (B)} {\it although cities may share administrative boundaries with others, they cluster with cities with no apparent geographical similarity}; and,
{\bf (C)} {\it there might be interesting correlations between urban and/or territorial indicators and the features extracted from the street-network topology of a given set of cities}.
The answering of such assumptions allows us to render better analysis of urban agglomerations by helping in the understanding of cities by comprehending how they are arranged within the geographical extent of their territorial boundaries.

Aiming to solve questions related to the urban scenario, a vast number of studies have been conducted to explain cities considering their intense flow of vehicles~\cite{Masucci2013:UrbanGrowth} and collective behaviors~\cite{Blumer1971:SocialProblems}, while others analyzed the accidents density in street networks~\cite{Anderson2009:RoadAccident} and the discrepancies between cities driven by their urban indicators~\cite{Grauwin2015:ComparativeScience}.
Furthermore, some authors investigated metrical and analytical methods applied to cities~\cite{Crucitti2006:SpatialCentrality,Costa2010:TransportationSystems}, others approached the assistance to the urban planning and design~\cite{Porta2009:StreetCentrality,Strano2012:GoverningEvolution,Spadon2017:UrbanInconsistencies}, and there are those who advanced with facility-location analysis and planning in street meshes~\cite{Li2016:SpatialOptimization}.
However, although cluster analysis has been less focused~\cite{Strano2013:StreetNetworks,Domingues2017:Topological}, it is still an important toolset~\cite{Pan2013:ProblemsSmartCities}.

Two state-of-the-art works used clustering techniques to analyze groups of cities, but both of them left open questions to be explored.
The first one had the intention to measure the similarity among ten European cities~\cite{Strano2013:StreetNetworks}, while the second one performed an eye-based cluster evaluation considering the proximity and overlap of 1,150 cities, mainly from the Anglo-Saxon America~\cite{Domingues2017:Topological}.
Their lack of proficiency is mainly because they do not employ clustering algorithms in the same fashion that we do, including validation metrics and analytical indicators.

In this paper, we contribute with a methodology that advances the analysis of cities modeled as complex networks.
To present our contributions, this paper is organized as follows:
Section~\ref{sect:methodology} displays our methodology while explaining the validation of its results;
Section~\ref{sect:results} discusses the results about the applicability of the proposed methods; and,
Section~\ref{sect:conclusion} presents the conclusions and final remarks.

\newpage
\section{Methodology}
\label{sect:methodology}

\begin{figure}[!htb]
    \centering
    \includegraphics[width=\linewidth]{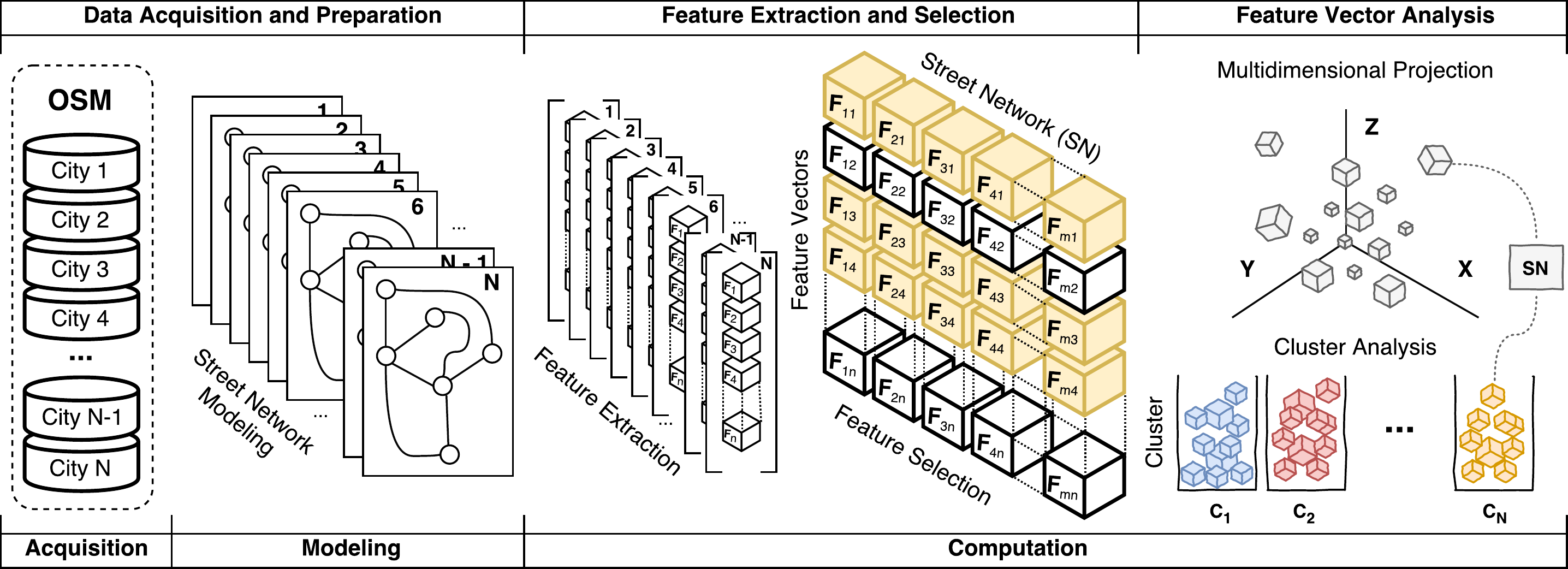}
    \caption
    {Methodology for street-network characterization through feature-vector and cluster analysis based on data {\bf A}cquisition, {\bf M}odeling, and {\bf C}omputation.
    The methodology starts by acquiring digital maps of cities from the OpenStreetMap (OSM), such maps are used for the modeling of complex networks.
    The resulting networks are used in the processes of extraction and selection of topological-features.
    These features are analyzed according to data-mining methods of multidimensional projection and cluster detection.}
    \label{fig:methodology}
\end{figure}

Our methodology is based on the intersection of methods of data {\bf A}cquisition, {\bf M}odeling, and {\bf C}omputation, and it follows a process flow depicted in Figure~\ref{fig:methodology}.

\subsection{Preliminaries}
\label{subs:preliminaries}

Hereinafter, we represent complex networks as distance-weighted directed graphs.
Notice that, despite different, complex networks and graphs are considered to be equivalent.
A graph $\mathbf{G} = \{V, E\}$ is composed of a set of $|V|$ nodes and a set of $|E|$ edges.
Furthermore, each edge $e \in E$ is known to be an ordered pair $\left<o, d \right>$, in which $o \in V$ is named {\it origin} and $d \in V$ is named {\it destination}, $o \neq d$.
We provided to the edges a double-precision floating-point weight $d_{od}$, which refers to the {\it great-circle distance} between node $o$ and node $d$.
The {\it great-circle distance} refers to the Euclidean distance between two points on the surface of a sphere; which in our case, the sphere is a projection of the~Earth.

\subsection{Data acquisition and preparation}
\label{subs:data}

For each one of the 645 cities from the Brazilian state of Sao Paulo, we got their administrative boundaries and indicators related to territorial extension and demography from the Brazilian Institute of Geography and Statistics (IBGE)\footnote{\url{www.ibge.gov.br}}.
The boundaries served as shapefiles to crop data obtained from OpenStreetMap (OSM)\footnote{\url{www.openstreetmap.org}}, which is an open data repository and a social network of collaborative street mapping.
The OSM's data describe real-world abstractions represented by georeferenced objects.
These objects are described by means of its relations, which, in turn, refer to the streets (edges) and crossings (nodes) of a city, which were turned into complex networks where the edges intersect only at the nodes.

\begin{figure}[!b]
    \centering
    \includegraphics[width=\linewidth]{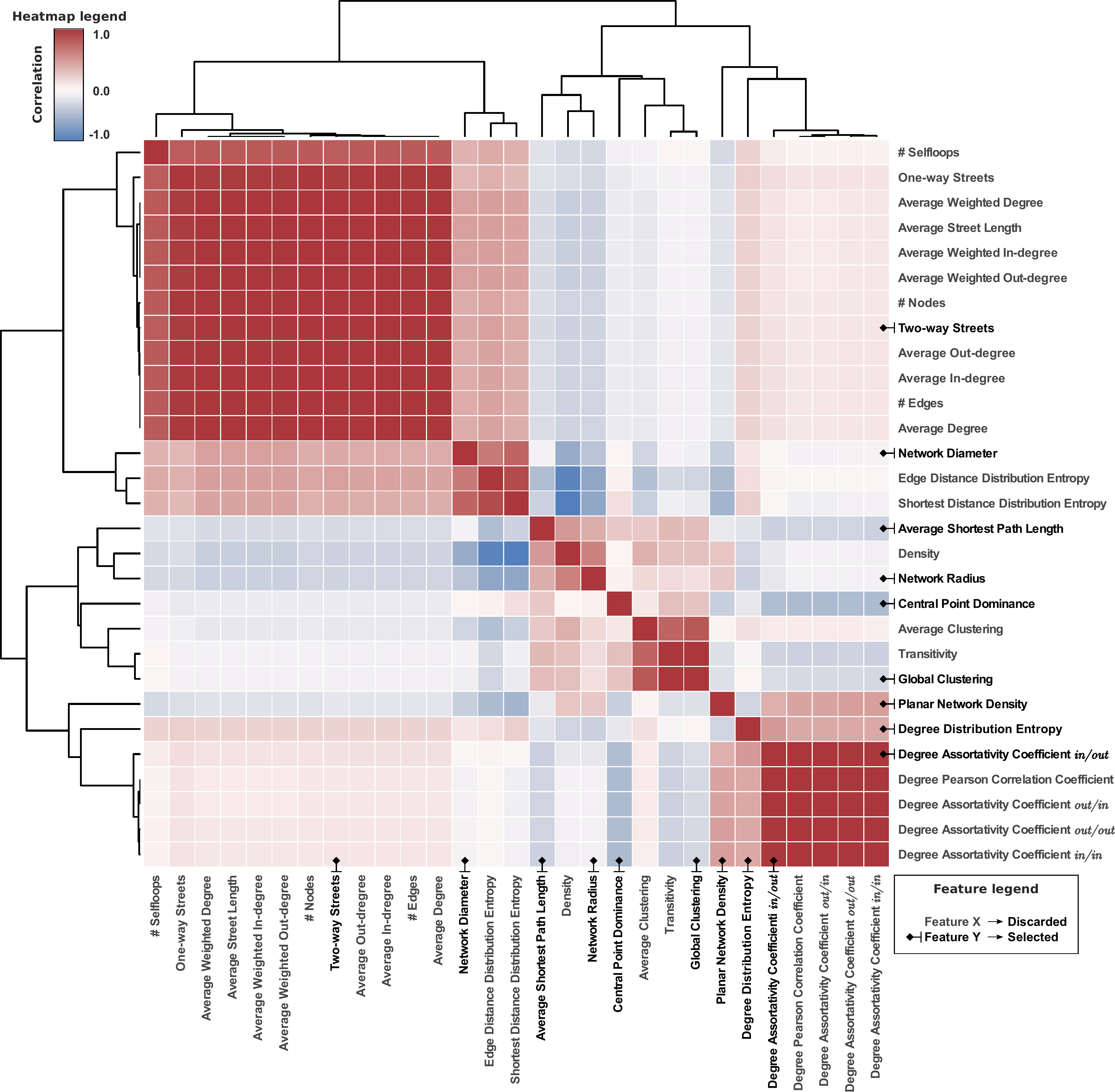}
    \caption
    {A visualization of the mutual-correlation matrix of all the metrics we considered.
The color describes the correlation between pairs of features.
The metrics were hierarchically grouped through a dendrogram by means of the correlation of their values.
Consequently, correlated metrics tend to stay in the same group; non-correlated metrics tend to be in separated groups.
Additionally, the metrics we selected are colored in black and highlighted by a diamond marker.}
    \label{fig:full-features}
\end{figure}

\subsection{Feature extraction and selection}
\label{subs:features}

Metrics of graphs, referred to as {\it features}, can be divided into local and global~\cite{Scripps2010:NetworkMining}; local metrics describe properties of individual elements that form the network, while global metrics characterize the whole network by a single value that is computed considering all of their elements.
We rather use global metrics than local ones because they allow straightforward comparison between~cities.

In order to gather these metrics, we designed a feature extractor to calculate a feature vector from any complex network given as input.
First, we selected various metrics as candidates to render characteristics about cities, from which 29 metrics were chosen by their potential in providing insights about a given street network (see Figure~\ref{fig:full-features} for details).
Such metrics were selected because they are linked to the network topology, which describes the streets of the cities.

After collecting all the metrics, we removed the non-relevant ones based on their mutual correlation.
We computed the Pearson correlation coefficient~\cite{Chiang2003:StatisticalAnalysis} for each pair of metrics; such coefficient is defined in the interval $[-1.0,1.0]$ where the extreme values indicate, respectively, the maximum negative and positive correlation, while $0.0$ indicates no linear correlation at all.
Following, we removed all the metrics with strong mutual correlation as indicated by the Pearson correlation in the interval $[-0.5,0.5]$.
In cases where any two metrics are outside this interval, one of the metrics was randomly discarded.
Such process of metrics selection ensures that just metrics that are unique and non-related with the others will be used to describe the cities.
Other processes of feature selection can be used in this step; even the multidimensional projection by itself can provide reasonable results.
Notice that, the reduction of the dimensionality of the data was not our main priority, but rather to find the most complete set of metrics, that is the one that better characterizes the networks, without including redundant information; and, to this end, features correlation plays an import role.
All metrics are depicted in Figure~\ref{fig:full-features}; the ones that remained, 9 out of 29, were highlighted and are defined according to Costa {\it et al.}~\cite{Costa2007:SurveyMeasurements}, as follows:

\paragraph{\it{\textbf{Degree Distribution Entropy}}} ($\mathcal{H}$).
The degree distribution of a network describes the probability of finding a vertex with a given degree.
Whereas, the entropy represents the amount of uncertainty and randomness in a certain piece of information.
By using the entropy in a city degree distribution, we can measure the uncertainty between street connections.
Equation~\ref{eq:entropia} describes such metric, where $P_k$ represents the ratio of nodes with degree $k$.

\paragraph{\it{\textbf{Average Shortest Path}}} ($\mathscr{L}$).
It quantifies the average of all shortest paths ($d^{S}_{ij}$) that link all the pairs of nodes in a complex network (Equation~\ref{eq:caminhos_minimos}), it is used to quantify the capacity of locomotion through the shortest paths of a city.

{\noindent%
\begin{minipage}{.45\textwidth}
    \centering
    \begin{equation}
        \mathcal{H} = - \sum_{k = 0}^{\infty}{P_k \times \log(P_k) }
        \label{eq:entropia}
    \end{equation}
\end{minipage}
\begin{minipage}{.45\textwidth}
    \centering
    \begin{equation}
        \mathscr{L}=\frac{\sum_{i=1}^{|V|}\sum_{j=1}^{|V|} d^{S}_{ij}}{|V|(|V|-1)}
        \label{eq:caminhos_minimos}
    \end{equation}
\end{minipage}}

\paragraph{\it{\textbf{Degree Assortativity Coefficient}}} ($\mathcal{R}$).
It refers to the {\it in and/or out} degree correlation between pairs of nodes.
That is, positive values indicate that nodes with similar degrees tend to connect to each other, while negative values indicate the same, but regarding nodes with different degrees.
It can be understood as the probability of moving from an unimportant street to an important one based only on the number of adjacent streets to both of them.
Equation~\ref{eq:assortativity} uses $e_{xy}$ to refer to the fraction of edges that join together vertices with degree $x$ and $y$, $a_x$ and $b_y$ to the fractions of edges that start and end at vertices with degree $x$ and $y$; and, $\sigma_a$ and $\sigma_b$ to the standard deviations of the distributions $a_x$ and $b_y$.

\paragraph{\it{\textbf{Eccentricity}}} ($\mathcal{E}$).
This metric is a local one, measuring for a vertex the longest shortest distance between all the other vertices of a given graph~\cite{Hage1995:Eccentricity} (see Equation~\ref{eq:eccentricity}).
In a global perspective, the greatest eccentricity from a graph is known to be the {\it network diameter}, while the smallest one is regarded as the {\it network radius}.
They can reveal cities that may suffer from access issues by being sparse if the radius of a network is too small when compared to its diameter.

{\noindent%
\begin{minipage}{.45\textwidth}
    \centering
    \begin{equation}
        \mathcal{R} = \frac{\sum_{xy} xy(e_{xy} - a_x b_y)}{\sigma_a \sigma_b}
        \label{eq:assortativity}
    \end{equation}
\end{minipage}
\begin{minipage}{.45\textwidth}
    \centering
    \begin{equation}
        \mathcal{E}_i = \frac{1}{max \{ d^{S}_{ij} | \forall j \in V \}}
        \label{eq:eccentricity}
    \end{equation}
\end{minipage}}

\paragraph{\it{\textbf{Planar Network Density}}} ($\mathcal{D}$).
The density of a planar graph is defined as the ratio between the number of edges $E$ and the number of all possible edges in a network with $N$ nodes with no intersecting edges.
It can be used to describe how dense is the street mesh of a city or a neighborhood.
The metric is unique to each network, once the position of the nodes interferes in the number of edges.
It is an algorithmic adaptation of the graph density~\cite{Brath2015:GraphAnalysis}, described in Equation~\ref{eq:densidade}.

\paragraph{\it{\textbf{Central Point Dominance}}} ($C^{P}_{D}$).
This metric assesses the global centrality of a whole network by means of its network's betweenness deviation, which is a distance-based centrality metric.
Values close to 0 indicate plenty of distance-efficient routes similar to the shortest one; whereas, values close to 1 indicate that the network might become vulnerable without its central node because the node might be used to connect different components, serving as an access point ({\it e.g.} bridges and tunnels).
In Equation~\ref{eq:central_point}, $\bar{v}$ is the node with the highest betweenness and $\mathcal{B}(v)$ is the normalized betweenness of the node $v$ that lies in the range $[0,1]$.

{\noindent%
\begin{minipage}{.45\textwidth}
    \centering
    \begin{equation}
        \mathcal{D} = \frac{|E|}{|N|\,(|N|-1)}
        \label{eq:densidade}
    \end{equation}
\end{minipage}
\begin{minipage}{.45\textwidth}
    \centering
    \begin{equation}
        C^{P}_{D} = \frac{\sum^{|V|}_{v} \mathcal{B}_{\bar{v}} - \mathcal{B}_{v}}{|V|(|V|-1)}
        \label{eq:central_point}
    \end{equation}
\end{minipage}}

\paragraph{\it{\textbf{Two-way Streets}} ($\mathcal{T}_w$).}
It refers to the number of double edges in a network, which are edges that provide two-way routes between the same pair of nodes.
This metric follows Equation~\ref{eq:global_clustering}, in which $f_{ij}$ is a clause-based auxiliary function.

\paragraph{\it{\textbf{Global Clustering}}} ($\mathcal{G}_c$).
The metric, which is described by Equation~\ref{eq:global_clustering}, consists of the {\it fraction} of the number of triangles $\mathbb{N}_\triangle$ and triples $\mathbb{N}_3$ of the network.
It refers to how the streets tend to cluster in the crossings of a given city, such that the greater the value the more possibilities of locomotion in fewer steps.

{\noindent%
\begin{minipage}{.60\textwidth}
    \centering
    \begin{equation}
        \mathcal{T}_w = \frac{\sum_{\left<i,j\right>}^{E} f_{ij}}{2}, \quad f_{ij} =
        \begin{cases}
            1, \quad &\left<j,i\right> \in E \\[1mm]
            0, \quad &otherwise
        \end{cases}
        \label{eq:global_clustering}
    \end{equation}
\end{minipage}
\begin{minipage}{.39\textwidth}
    \centering
    \begin{equation}
        \mathcal{G}_c = \frac{(3 \times \mathbb{N}_\triangle)}{\mathbb{N}_3}
        \label{eq:global_clustering}
    \end{equation}
\end{minipage}}

\subsection{Feature vector analysis}
\label{subs:mining}

In this step, we focused on two methods from the data mining literature, the first one of multidimensional projection and the second one of clustering detection.
Multidimensional projection allows the visualization of data by reducing its dimensional space, revealing particularities and behaviors to be explored through cluster-based analysis.
Cluster analysis, in turn, focuses on the study of data interactions, inferring that two elements are similar because they are in the same cluster or dissimilar because they are in different ones.
Consequently, the combination of these two methods contributes to the assessment of cities by their potential to reveal patterns that are not evident through an eye-based analysis.

Regarding multidimensional projection, our methodology consists of using two techniques~\cite{Spiwok2015:Folding}; the first one is named Isomap and the second one is known as Principal Components Analysis (PCA).
Isomap is a nonlinear dimensionality reduction technique, which provides an embedding in a lower dimension while maintaining the geodesic distance between the data elements.
Contrarily, PCA is a linear technique, which uses orthogonal conversions to turn a set of variables into linearly uncorrelated values with the largest possible variance.
To choose both techniques, we used knowledge about the domain; we have kept track of some already-known dissimilar cities, seeking for approaches to distinguish them.

In the cluster analysis part, we used the technique KMeans~\cite{MacQueen1967:KMeans}, which splits the data into groups of equal variance, minimizing the sum-of-squares distance within clusters.
The KMeans algorithm assumes that (i) the distribution of features within each cluster resembles spheres, which means that all features have equal variance and they are independent of each other; (ii) regarding the cluster size, the dataset is balanced; and, (iii) the density of the clusters is similar.
The dataset we used consists of uncorrelated values and balanced instances of feature vectors, all of which have quasi-equal variance, meeting the algorithm requirements.
In addition, KMeans is widely used in the related literature due to its robustness, versatility, and scalability.
To validate our results we considered cluster quality metrics~\cite{Kremer2011:EvaluationMeasure}.
Their focus is to analyze the similarity between elements that have been assigned to the same cluster.
We used a combination of the Silhouette score~\cite{Rousseeuw1987:Silhouette} and the Dunn index~\cite{Dunn1974:DunnIndex}; both of which are known to be internal-quality metrics, not requiring a pre-labeled dataset.
The Silhouette is defined between $ [-1, 1] $ for each cluster, the closer to $1$ the better; it measures the cohesion and separation of clusters by evaluating how similar an element is in its own cluster when contrasted to other clusters.
To further enhance the reliability of our analysis, we applied the Dunn index, which is a cluster distance-based quality metric that measures the separation among clusters, whose values are in between of $ [0, \infty[ $.
In cases when the Dunn's index distance is greater than one, there is little or none cluster overlapping.
Using both together, we have a double validation of quality by means of cohesion and separation of our set of clusters.

\section{Results}
\label{sect:results}

\subsection{Relationship of population size and topological features}
\label{subs:projection}

\begin{wrapfigure}{R}{0.615\textwidth}
    \centering
    \includegraphics[width=\linewidth]{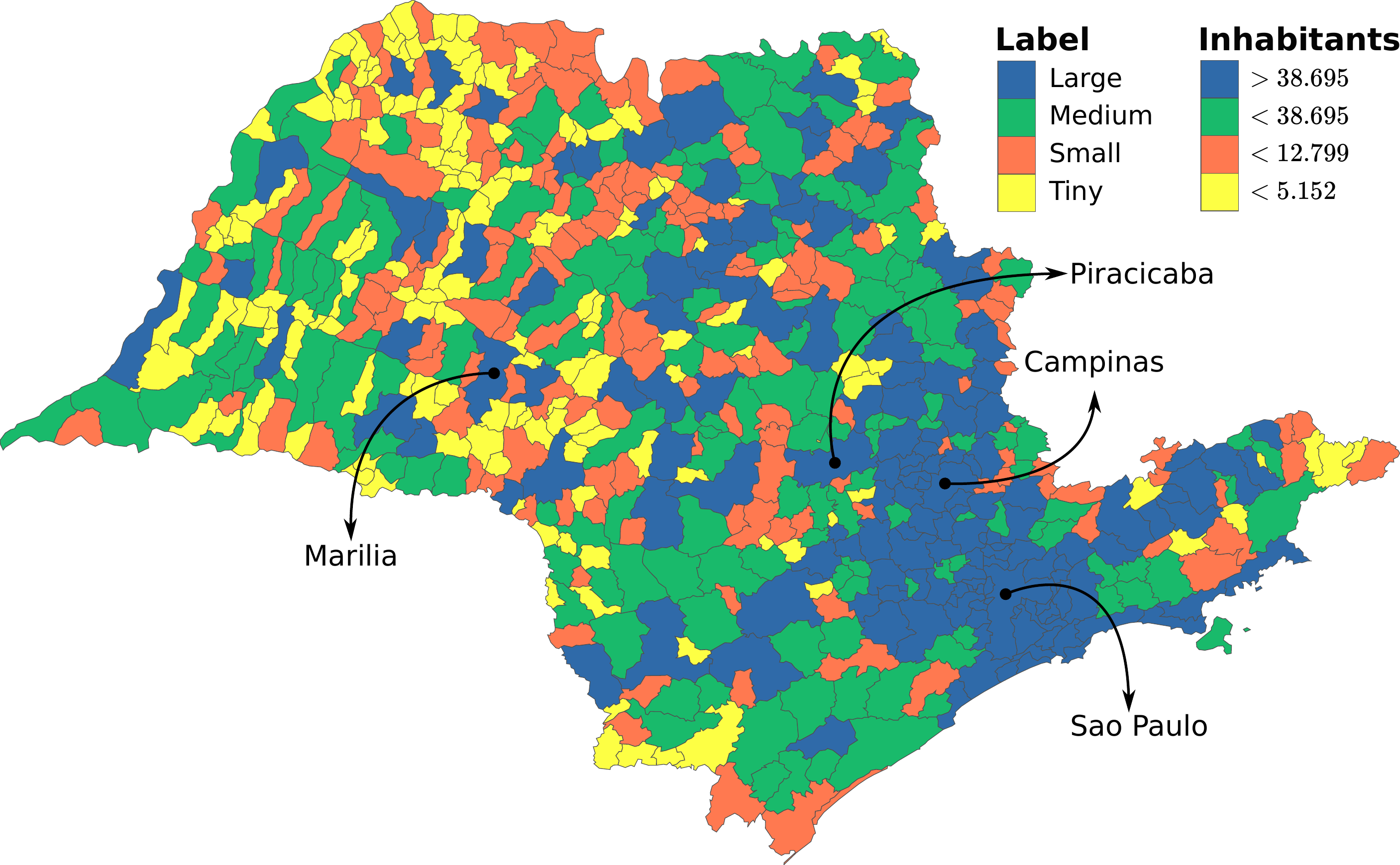}
    \caption
    {Urban indicator related to the population density of the cities of the state of Sao Paulo.
    The cities were divided into four classes that describe the number of inhabitants of each one of them.}
    \label{fig:population_density}
\end{wrapfigure}

\begin{figure}[!b]
    \centering
    \includegraphics[width=\linewidth]{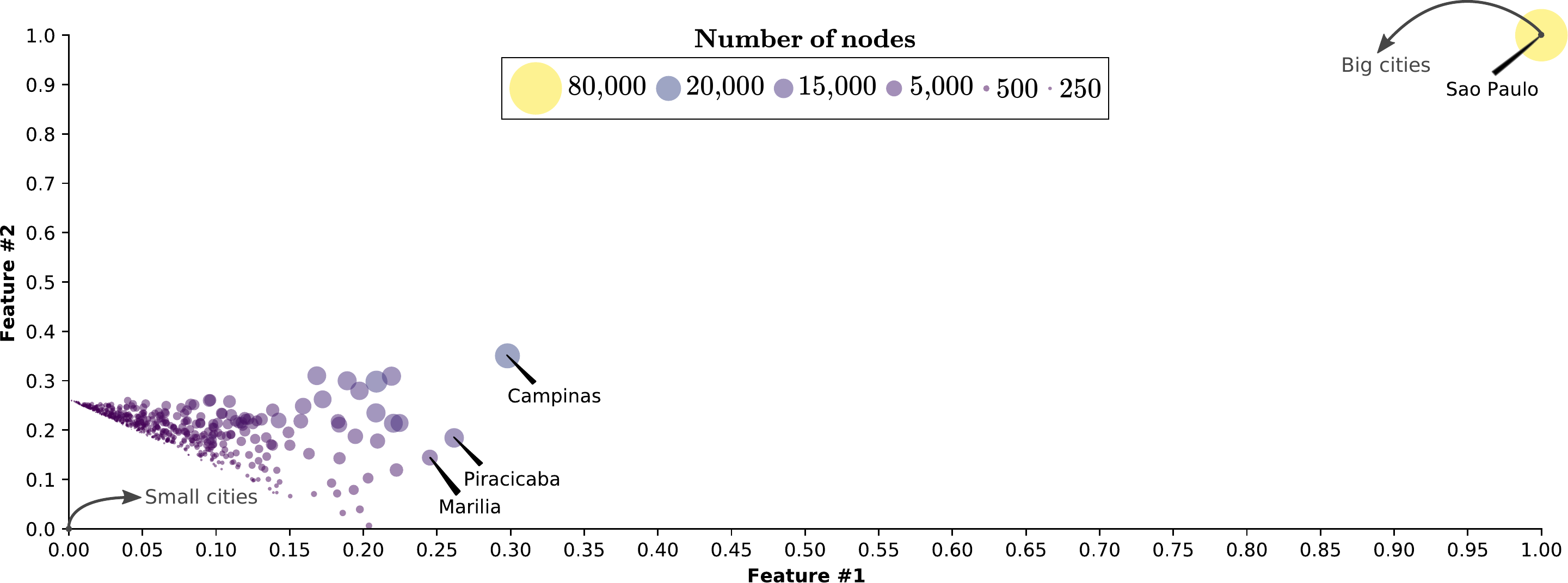}
    \caption
    {Projection of feature vectors in two dimensions by using PCA; the size of the points refers to the number of nodes (intersections) in the cities’ complex-network.
    The projected features reveal that the city of Sao Paulo (on the right-hand side) is an outlier when compared to the others (on the left-hand side).}
    \label{fig:features_size}
\end{figure}

With regard to the population density --- see Figure~\ref{fig:population_density} for details ---, the majority of the cities in our dataset is of tiny or small size, but the dataset has a substantial number of medium-sized cities and a small number of large-sized ones, including Sao Paulo --- the biggest~Brazilian city.
Prior analyses can be done by observing Figure~\ref{fig:features_size}, where cities (depicted as points) were sized by their number of nodes.

\begin{figure}[!t]
    \centering
    \includegraphics[width=\linewidth]{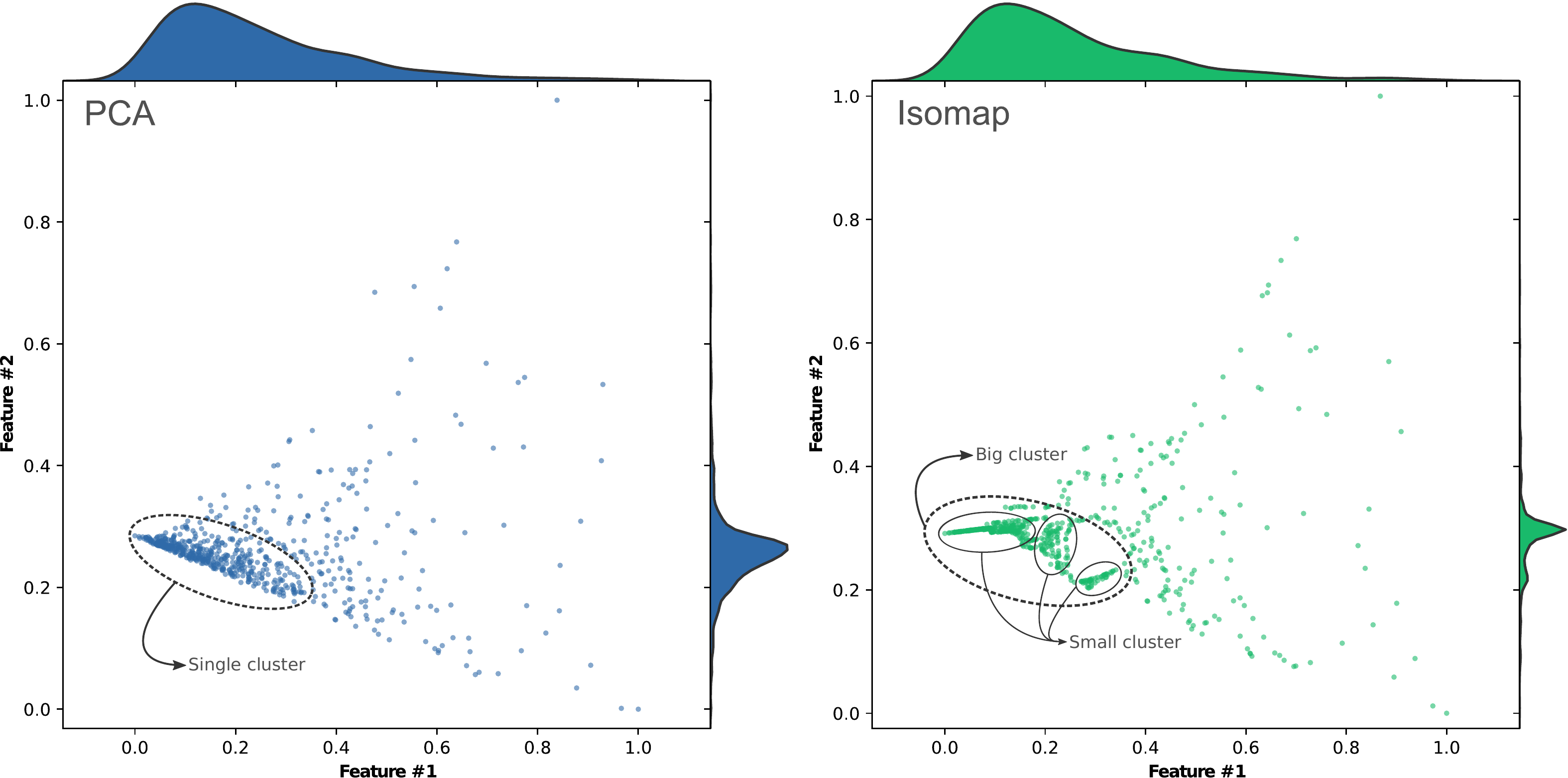}
    \caption
    {Projection of the features, excluding Sao Paulo, using PCA and Isomap.
    PCA shows a single dense area with many sparse data, while Isomap shows multiple dense areas together with several sparse data.
    As a consequence, PCA implies a single cluster while Isomap points to an inherent hierarchy of clusters.}
    \label{fig:pca_isomap}
\end{figure}

A first evidence that the topological features we selected can describe relevant knowledge about cities is the fact that Sao Paulo is isolated from the other ones in the PCA projection.
A similar fact can be observed on a small scale considering the large-sized city of Campinas and the medium-sized cities of Marilia and Piracicaba, which are apart from the main group of cities located on the left part of the image.
We believe that such behavior is connected to the demographics of the cities.
On a large scale, topological features can predict demographic characteristics of a city, whereas, on a small scale, they can reflect the neighborhoods that are densely or sparsely populated.
For a less unbalanced view, we removed Sao Paulo from the dataset, depicting in Figure~\ref{fig:pca_isomap} the normalized values of the feature vectors of the cities that remained using both PCA and Isomap techniques.

\begin{figure}[!b]
    \centering
    \includegraphics[width=\linewidth]{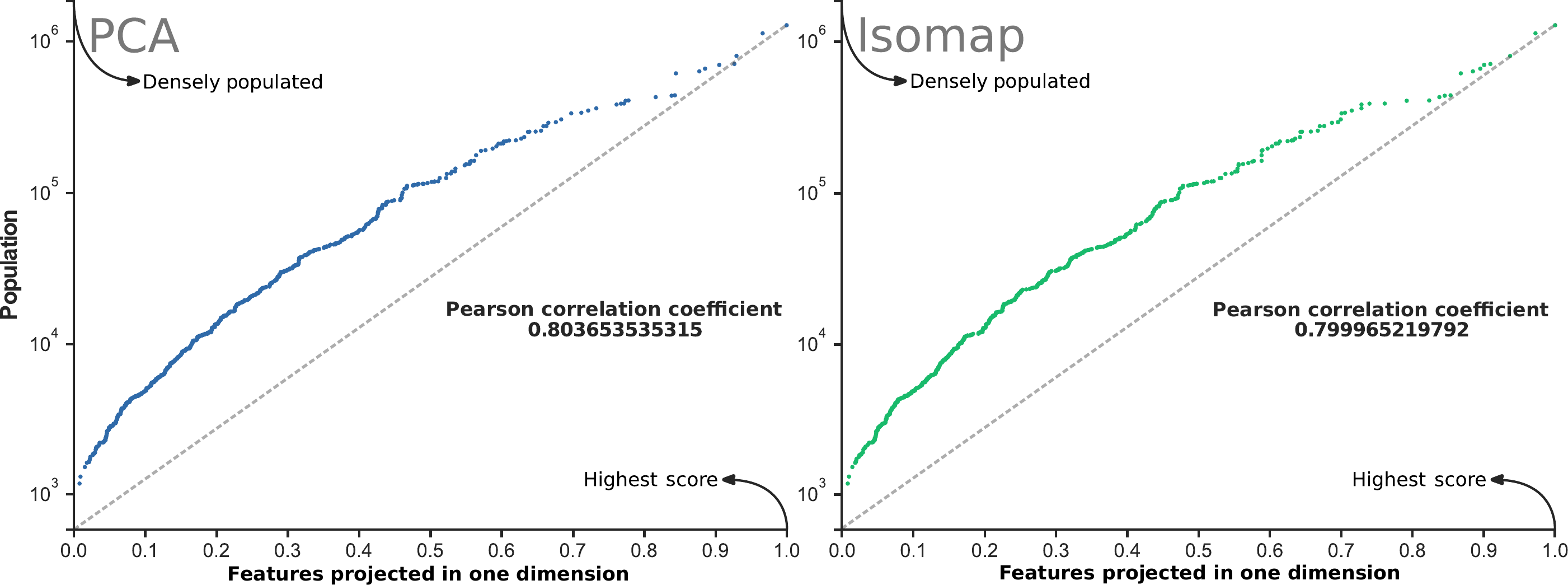}
    \caption
    {Correlation test between the population density and a one-dimension projection of the cities' topological-features regarding both PCA and Isomap.
    Both images show a strong correlation, revealing that, on a large scale, the topological features of the cities can indicate, or even predict, their demography.}
    \label{fig:population_relation}
\end{figure}

The two techniques show us that the majority of the data is concentrated in a small region, while the rest of it is sparse and distributed along the axes.
The main difference between both of them is that Isomap implies multiple areas with considerable density, while PCA has a single dense area and many sparse data.
This is evidence that tiny and small-sized cities tend to cluster isolating medium and large-sized cities that are too different from them.
Despite the fact that such cities tend to cluster, Isomap shows that they have particularities that make them split into smaller clusters inside a bigger one.
Also, it is safe to infer that by being scattered, medium-sized and large-sized cities have no clear pattern, but still, they may share common characteristics to be further explored with clustering algorithms.
Even so, we can show, by using correlation, that the network's demography can be inferred from the city's topology -- see~Figure~\ref{fig:population_relation}.

To prove that the network's demography can be inferred from the city's topology, we measured the relationship between the topological features and the demography by means of correlation.
To this end, we reduced the dimensionality of the feature vectors of each city to one, using both techniques, PCA and Isomap, resulting in one single value for each one of the 645 cities.
Next, we correlated such values with the size of their population.
As a result, we got 0.803 and 0.799 of correlation for PCA and Isomap, respectively.
Both values indicate that the data has a strong correlation, allowing us to state that in the case of the Brazilian state of Sao Paulo, topological features and demographics are strongly correlated.
Such pattern opens doors for new investigations, as the ones placed by the dynamics of the social behavior; as in the case of criminality and mobility.

\subsection{Relationship of cluster assignment and territorial extension}
\label{subs:clustering}

The cluster analysis aimed at the identification of the best number of clusters to describe our dataset.
Consequently, we exhaustively tested the KMeans' cluster-quantity parameter from 2 to 644 clusters --- the total number of cities without considering Sao Paulo.
During the test, {\it we were seeking for the greatest average Silhouette score (AVG) only when the Dunn index (DNN) was larger than one}.

\begin{figure}[!htb]
    \centering
    \includegraphics[width=\linewidth]{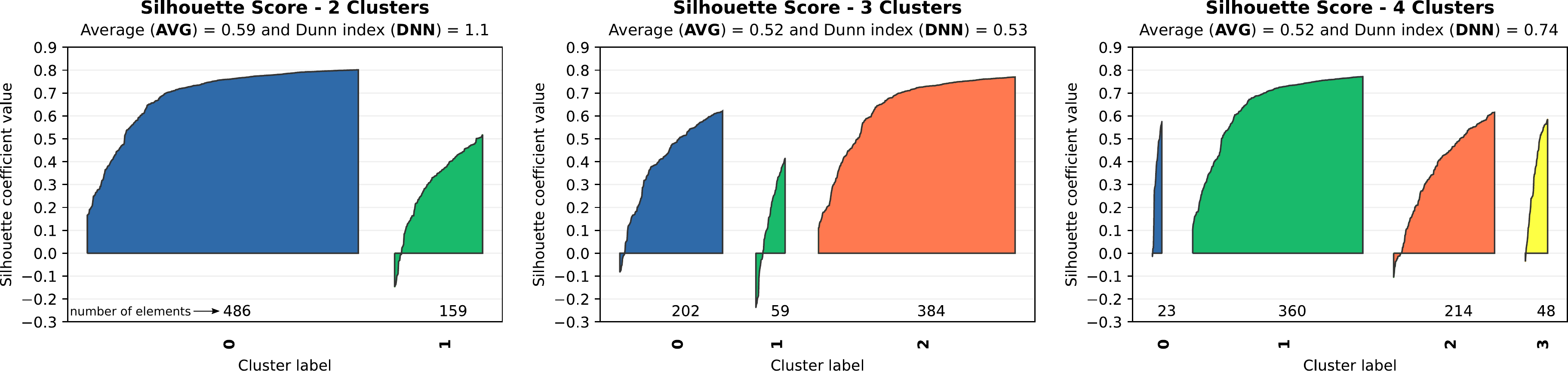}
    \caption
    {Silhouette analysis of the subset of our data without Sao Paulo, in which clusters are represented as color-coded polygons.
    In each scenario that we have tested, the results were validated according to the Dunn index together with the Silhouette score.
    Although we have depicted the first three tested scenarios, which are also the best ones, the experiment considers a total of 643 scenarios.}
    \label{fig:silhouette-sp}
\end{figure}

The previous experiment suggests that the best way to split our data is into two clusters.
Such configuration has an AVG of 0.59 and a DNN of 1.10 (see Figure~\ref{fig:silhouette-sp}).
When dividing the data into two, the clusters are better balanced rather than when considering Sao Paulo --- a big outlier --- as part of the dataset.

\begin{figure}[!t]
    \centering
    \begin{subfigure}[t]{0.5\textwidth}
        \centering
        \includegraphics[width=\linewidth]{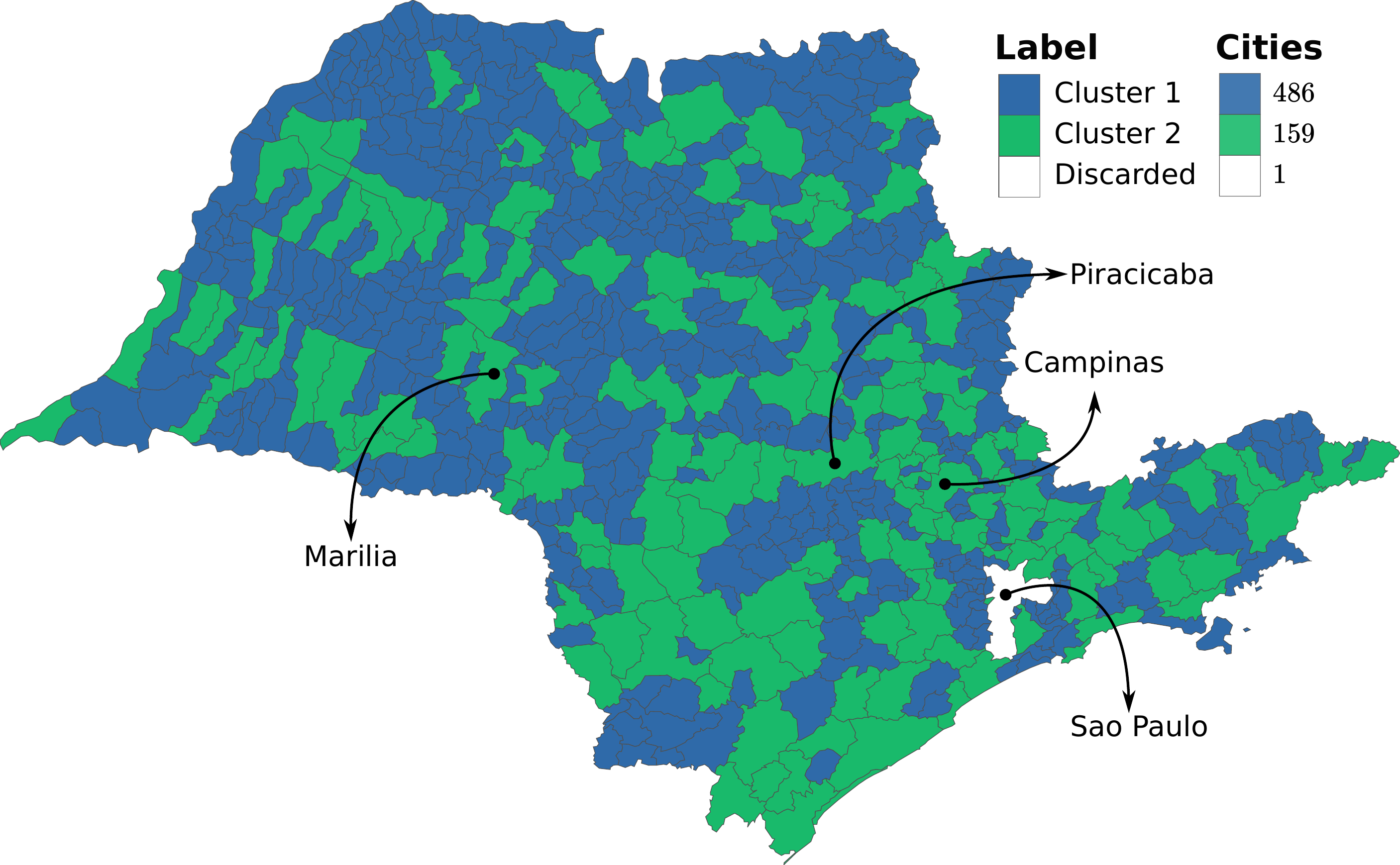}
        \caption{Clustering Sao Paulo's cities.}
        \label{fig:second_cluster}
    \end{subfigure}%
    ~
    \begin{subfigure}[t]{0.5\textwidth}
        \centering
        \includegraphics[width=\linewidth]{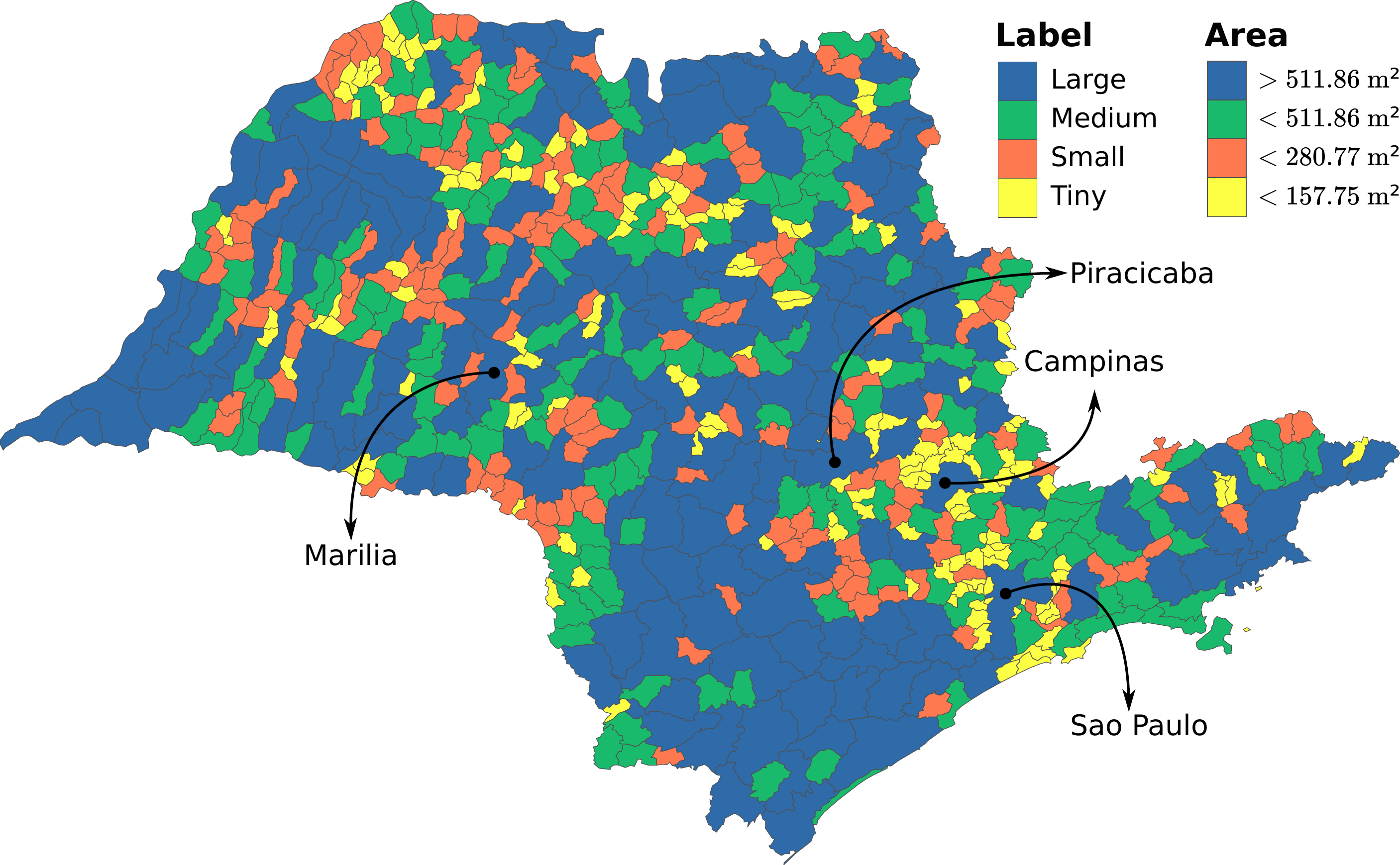}
        \caption{Sao Paulo's territorial extension.}
        \label{fig:administrative_boundaries}
    \end{subfigure}%
    \caption
    {Investigating cities through clustering techniques; Figure~\ref{fig:second_cluster} shows the results of the clustering of topological features when removing the Sao Paulo city from the dataset; this layout has an average Silhouette of 0.59 and Dunn Index of 1.1.
    Figure~\ref{fig:administrative_boundaries} describe the area within the cities' administrative boundaries.}
\end{figure}

Subsequently, we investigated the reason why the cities were better arranged into only two clusters.
By analyzing indicators related to population and territory extension, we found that 61.20\% of the state's population is in the first cluster and 38.80\% is in the second one (see Figure~\ref{fig:second_cluster}), and that the first cluster is mainly populated with cities that are considered to be of tiny or small territorial extension (see Figure~\ref{fig:administrative_boundaries}), while the second cluster has the opposite behavior.
Bearing in mind that our dataset does not imply any relationship between indicators of territorial extension and population density, we concluded that the relation that favored two clusters, as the arrangement with best values of Silhouette and Dunn index, was the territorial extension of the cities.
Hence, we found evidence that there is a significant relationship between topological features, territorial extension, and demographics of the cities of Sao Paulo state.

The relationship between the cluster arrangement and territorial extension can be understood as the way cities organize within their available space.
In fact, regarding the territorial extension, 30.51\% of the cities from the first cluster are tiny-sized, 31.13\% are small-sized, 25.78\% are medium-sized, and 12.58\% are large-sized; whereas, 7.59\% of the ones from the second cluster are tiny-sized, 6.32\% are small-sized, 22.78\% are medium-sized, and 63.29\% are large-sized.
Therefore, cities in the first cluster can be considered smaller and heavily populated, while the ones in the second cluster are larger and less populated.

\subsection{Discussions on results generalization}
We have chosen to present a joint of direct findings and analytical conclusions in our results section.
This was done so that one can follow the practical application of the proposed methodology in a way that can be adapted and generalized for different domains and scenarios.
Our methodology can also be used in non-urban applications, such as in the characterization of the topology in any group of complex networks, however, depending on the specificities of the domain, it may be necessary to use different network metrics and features to be more effective.

Additionally, while our findings cannot be generalized for any set of cities, we believe that the proposed methodology can be used to find non-trivial properties in different urban scenarios and not only to the cities that shape the state of Sao Paulo.
Comprising a straightforward framework of analysis that can be useful to the academic community and cities' governing body, {\it e.g.} planners and designers.

Finally, our proposal has intricacies that can be explored in further studies:
{\bf (1)} using hierarchical clustering to reveal additional knowledge, which may also demand prior expertise about the cities ({\it e.g.} history and geography); and,
{\bf (2)} using more complex feature selection techniques such as fractal-dimension based methods or by applying ones related to mutual information.
Notice that, this refinement might reveal other patterns of the data, but will not change the ones we discussed; and,
{\bf (3)} including non-topological features to capture different characteristics of cities, enhancing our methods capabilities and its versatility.

\section{Conclusion}
\label{sect:conclusion}

In this paper we proposed a three-folded method encompassing the data {\bf A}cquisition, {\bf M}odeling, and {\bf C}omputation.
Furthermore, our methodology comprises the following phases: {\it Data Acquisition and Preparation}, {\it Feature Extraction and Selection}, and {\it Feature Vector Analysis}; culminating in the use of multidimensional projection and cluster analysis algorithms to assess feature vectors of complex-network metrics.
To validate our proposal, we investigated the following hypotheses:
{\bf (A)} {\it the network topology is a tool-set that can reveal groups of cities with similar characteristics, potentially revealing disparities};
{\bf (B)} {\it although cities may share administrative boundaries with others, they cluster with cities with no apparent geographical similarity}; and,
{\bf (C)} {\it there might be interesting correlations between urban and/or territorial indicators and the features extracted from the street-network topology of a given set of cities}.
Such hypotheses were investigated by analyzing relations between 645 cities that constitute the Brazilian state of Sao Paulo.
Our main findings confirm the hypotheses of our work, allowing us to state that, on a large scale, the topological features of the cities can indicate, or even predict, their demography and that cities group themselves by means of their territorial extension, which describes the way that cities organize within their available space.
Therefore, our main contributions are:
{\bf (i)  } {\it the description of how the network topology is capable of revealing groups of cities with similar characteristics};
{\bf (ii) } {\it the correlation analysis between the demography of the cities and their features}; and,
{\bf (iii)} {\it the discussion of why cities cluster with other cities distant apart instead of with those that they share boundaries with}.
As a future work, we will measure the similarity between cities by means of non-topological features, looking for discrepancies in the collective behavior that emerges from this same set of cities.

\section*{Acknowledgment}
\label{sect:acknowledgement}

We are thankful to CNPq (grant 167967/2017-7), FAPESP (grants 2014/25337-0, 2016/17078-0 and 2017/08376-0) and CAPES that supported this research.

\begin{small}

\end{small}


\begin{thebibliography}{10}
    \bibitem{Boccaletti2006:StructureDynamics}
    Boccaletti, S; Latora, V; Moreno, Y; Chavez, M; Hwang, D:
    \newblock {Complex networks: Structure and dynamics}.
    \newblock Physics Reports \textbf{424}(4-5) (feb 2006) 175--308

    \bibitem{Masucci2013:UrbanGrowth}
    Masucci, AP; Stanilov, K; Batty, M:
    \newblock {Limited Urban Growth: London's Street Network Dynamics since the
     18th Century}.
    \newblock {PLoS} {ONE} \textbf{8}(8) (aug 2013)

    \bibitem{Blumer1971:SocialProblems}
    Blumer, H:
    \newblock Social problems as collective behavior.
    \newblock Soc Probl \textbf{18}(3)(1971) 298--306

    \bibitem{Anderson2009:RoadAccident}
    Anderson, TK:
    \newblock Kernel density estimation and k-means clustering to profile road
     accident hotspots.
    \newblock Accident Analysis \& Prevention \textbf{41}(3) (2009) 359--364

    \bibitem{Grauwin2015:ComparativeScience}
    Grauwin, S; Sobolevsky, S; Moritz, S; G{\'o}dor, I; Ratti, C:
    \newblock Towards a comparative science of cities: Using mobile traffic records
     in new york, london, and hong kong.
    \newblock In: Computational approaches for urban environments.
    \newblock Springer (2015) 363--387

    \bibitem{Crucitti2006:SpatialCentrality}
    Crucitti, P; Latora, V; Porta, S:
    \newblock {Centrality measures in spatial networks of urban streets}.
    \newblock Phys Rev E: Stat Nonlinear Soft Matter Phys \textbf{73}(3) (mar 2006)

    \bibitem{Costa2010:TransportationSystems}
    Costa, LF; Traven{\c{c}}olo, BAN; Viana, MP; Strano, E:
    \newblock {On the efficiency of transportation systems in large cities}.
    \newblock {EPL} (Europhysics Letters) \textbf{91}(1) (jul 2010)

    \bibitem{Porta2009:StreetCentrality}
    Porta, S; Latora, V; Wang, F; Strano, E; Cardillo, A; Scellato, S;
     Iacoviello, V; Messora, R:
    \newblock {Street Centrality and Densities of Retail and Services in Bologna,
     Italy}.
    \newblock Environment and Planning B: Planning and Design \textbf{36}(3) (2009)
     450--465

    \bibitem{Strano2012:GoverningEvolution}
    Strano, E; Nicosia, V; Latora, V; Porta, S; Barth{\'{e}}lemy, M:
    \newblock {Elementary processes governing the evolution of road networks}.
    \newblock Scientific Reports \textbf{2} (mar 2012)

    \bibitem{Spadon2017:UrbanInconsistencies}
    Spadon, G; Gimenes, G; Rodrigues-Jr, JF:
    \newblock {Identifying Urban Inconsistencies via Street Networks}.
    \newblock Volume 108; Elsevier BV (2017) 18 -- 27 International Conference on
     Computational Science, ICCS 2017, 12-14 June 2017, Zurich, Switzerland.

    \bibitem{Li2016:SpatialOptimization}
    Li, X; Parrott, L:
    \newblock An improved genetic algorithm for spatial optimization of
     multi-objective and multi-site land use allocation.
    \newblock Comput Environ Urban Syst~\textbf{59}(2016)

    \bibitem{Strano2013:StreetNetworks}
    Strano, E; Viana, M; Costa, LF; Cardillo, A; Porta, S; Latora, V:
    \newblock {Urban Street Networks, a Comparative Analysis of Ten European
     Cities}.
    \newblock Environment and Planning B: Planning and Design \textbf{40}(6) (dec
     2013) 1071--1086

    \bibitem{Domingues2017:Topological}
    Domingues, GS; Silva, FN; Comin, CH; Costa, LF:
    \newblock Topological characterization of world cities.
    \newblock arXiv preprint arXiv:1709.08244 (2017)

    \bibitem{Pan2013:ProblemsSmartCities}
    Pan, G; Qi, G; Zhang, W; Li, S; Wu, Z; Yang, LT:
    \newblock Trace analysis and mining for smart cities: issues, methods, and
     applications.
    \newblock IEEE Commun Mag \textbf{51}(6) (2013)

    \bibitem{Scripps2010:NetworkMining}
    Scripps, J; Nussbaum, R; Tan, PN; Esfahanian, AH:
    \newblock {Link-Based Network Mining}.
    \newblock In: Structural Analysis of Complex Networks.
    \newblock Springer Nature (sep 2010) 403--419

    \bibitem{Chiang2003:StatisticalAnalysis}
    Chiang, C:
    \newblock Statistical Methods of Analysis.
    \newblock World Scientific (2003)

    \bibitem{Costa2007:SurveyMeasurements}
    Costa, LF; Rodrigues, FA; Travieso, G; Boas, PRV:
    \newblock Characterization of complex networks: A survey of measurements.
    \newblock Adv Phys \textbf{56}(1) (2007) 167--242

    \bibitem{Hage1995:Eccentricity}
    Hage, P; Harary, F:
    \newblock Eccentricity and centrality in networks.
    \newblock Social networks, \textbf{17}(1), (1995), 57-63.
    
    \bibitem{Brath2015:GraphAnalysis}
    Brath, R; and Jonker, D:
    \newblock Graph Analysis and Visualization: Discovering Business Opportunity in Linked Data.
    \newblock John Wiley \& Sons (2015).

    \bibitem{Spiwok2015:Folding}
    Spiwok, V; Oborský, P; Pazúriková, J; Křenek, A; Králová, B:
    \newblock Nonlinear vs. linear biasing in trp-cage folding simulations.
    \newblock J Chem Phys \textbf{142}(11) (2015)

    \bibitem{MacQueen1967:KMeans}
    MacQueen, J; et~al:
    \newblock Some methods for classification and analysis of multivariate
     observations.
    \newblock In: Proceedings of the fifth Berkeley symposium on mathematical
     statistics and probability. Volume~1. (1967) 281--297

    \bibitem{Kremer2011:EvaluationMeasure}
    Kremer, H; Kranen, P; Jansen, T; Seidl, T; Bifet, A; Holmes, G;
     Pfahringer, B:
    \newblock An effective evaluation measure for clustering on evolving data
     streams.
    \newblock In: Proceedings of the 17th ACM SIGKDD International Conference on
     Knowledge Discovery and Data Mining. KDD '11, New York, NY, USA, ACM (2011)
     868--876

    \bibitem{Rousseeuw1987:Silhouette}
    Rousseeuw, Peter J Silhouettes:
    \newblock A graphical aid to the interpretation and validation of cluster analysis.
    \newblock J Comput Appl Math, \textbf{20}(1), (1987), p. 53-65.
     
    \bibitem{Dunn1974:DunnIndex}
    Dunn, Joseph C:
    \newblock Well-separated clusters and optimal fuzzy partitions.
    \newblock Journal of cybernetics \textbf{142}(1), (1974), 95-104.
\end{thebibliography}
\end{document}